\documentclass{mem}
\usepackage{natbib}\usepackage{txfonts}\usepackage{balance}
\usepackage{graphicx}
\usepackage[a4paper,breaklinks,dvipdfm]{hyperref}
\idline{75}{282}
\begin{document}
\def\teff{$T\rm_{eff }$}
\def\kms{$\mathrm {km s}^{-1}$}

\title{
Blue straggler stars in Globular Clusters: chemical and kinematical properties.
}

   \subtitle{}

\author{
L. \,Lovisi\inst{1}, 
A. \, Mucciarelli\inst{1},
B. \, Lanzoni\inst{1},
F. R. \, Ferraro\inst{1},
E. \, Dalessandro\inst{1}
         }

  \offprints{L. Lovisi}

\institute{
Dipartimento di Fisica e Astronomia, Universit\`a degli Studi
di Bologna, Viale Berti Pichat 6/2, I--40127 Bologna, Italy\\
\email{loredana.lovisi2@unibo.it}
}

\authorrunning{Lovisi }

\titlerunning{Blue straggler stars in GCs}

\abstract{
Blue Straggler stars are present in all the properly observed Globular Clusters. 
They mimic a rejuvenated stellar population and their existence has been a puzzle for many years.
We performed an extensive spectroscopic survey of the surface abundances and the rotational 
velocities of Blue Straggler stars in different Globular Clusters, by using the high-­resolution 
spectrograph FLAMES@VLT in the UVES+GIRAFFE combined mode. In this contribution we show our main 
results obtained for M4, $\omega$ Centauri and M30.
}

\maketitle{}

\section{Introduction}
Blue Straggler stars (BSSs) are commonly defined as stars brighter and bluer (hotter) than the main 
sequence (MS) turnoff (TO). Firstly discovered in 1953 in the Globular Cluster (GC) M3 
(\citealt{Sandage53}), they lie along an extension of the MS in the color-magnitude 
diagram (CMD). According to their position in the CMD and also from direct measurements 
(\citealt{Shara97}), they are more massive than normal MS stars thus indicating
that a process able to increase the initial mass of a single star must be at work. 
Two main scenarios have been proposed for their formation: BSSs could be the end-products 
of stellar mergers induced by collisions (COL-BSSs; \citealt{Hills76}), or they may form by the
mass-transfer activity between two companions in a binary system (MT-BSSs; \citealt{McCrea64}), 
possibly up to the complete coalescence of the two stars.
Hence, BSSs represent the link between standard stellar evolution and cluster 
dynamics: they are able to give information about the dynamical history of the cluster \citep{Ferraro12}, the 
role of the dynamics on stellar evolution, the amount of binary sistems and the role of binaries in the 
cluster evolution. Indeed, distinguishing COL-BSSs from MT-BSSs is crucial to use these stars 
as GC dynamical probes.\\ 
The spectroscopic analysis seems to be the most promising way to discriminate 
between the two formation channels: in fact, it allows to derive both chemical abundances and rotational velocities. 
From a chemical point of view, COL-BSSs are not expected to show any abundance anomaly, since hydrodynamic simulations 
suggest that very little mixing should occur between their inner cores and their outer envelopes \citep{Lombardi95}.
At odds, depleted surface abundances of C and O are expected for MT-BSSs, since the accreted material should come 
from the core region of a peeled parent star where nuclear processing has occurred (\citealt{Sarna96}.
Concerning the rotational velocities, the theoretical scenario is more complicated. 
MT-BSSs are expected to have high rotational velocities because of the angular momentum transfer with the mass \citep{Sarna96}.
Unfortunately, accurate simulations are lacking, mostly because of the difficulty in following the evolution of a 
hydrodynamic system (the mass transfer between binary components) for the lenght of time required for the system to merge.
According to some authors \citep{Benz87}, also COL-BSSs should rotate fast. Nevertheless, braking mechanisms (like magnetic braking 
and disk locking) may intervene (both for MT- and COL-BSSs) with efficiencies and time-scales not well known yet 
(see \citealt{LeonardLivio95} and \citealt{Sills05}), thus preventing a clear prediction of the expected rotational
velocities.\\
In this context the advent of 8-10 meters class telescopes equipped with multiplexing capability spectrographs 
has given a new impulse to the study of the BSS properties.\\ 
By using the multi-object spectrograph FLAMES at ESO VLT in the UVES+GIRAFFE 
combined mode, an extensive survey has been performed in order to obtain abundance patterns and rotational velocities for 
representative samples of BSSs in a number of GCs. The first result has been obtained by \citet[] [hereafter F06]{Ferraro06a}: 
by measuring the surface abundance patterns of 43 BSSs in 47 Tuc, a sub-population of 6 BSSs with
a significant depletion of C and O with respect to the dominant population, has been discovered. 
This evidence has been interpreted as the presence of CNO burning products on the BSS 
surface and it is the first detection of a chemical  signature pointing to the MT formation process for BSSs in a GC.\\
The observations in 47 Tuc have also shown that most of the BSSs are slow rotators
(v $\sin(i) < $  10 \kms), with velocities compatible with those measured in TO stars (\citealt{LucatelloGratton03}). 
Only one BSS having a really large rotational velocity (v $\sin(i) \sim 80$ \kms) has been observed. 
No correlation has been found between CO depletion and rapid rotation.
\section{M4}
\begin{figure}[]
\resizebox{\hsize}{!}{\includegraphics[clip=true]{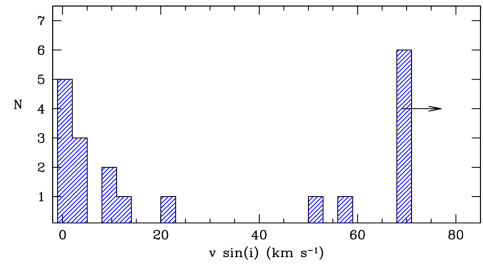}}
\caption{Rotational velocity distribution for BSSs in M4. Upper limits are marked with an arrow.
\footnotesize
}
\label{rotm4}
\end{figure}

M4 is the closest Galactic GC (2.1 kpc, \citealt{Harris96}).
We observed 20 BSSs and 53 TO stars along the entire extension of the cluster. We found \citep{Lovisi10} that the C and O abundances of the BSSs 
are totally in agreement with those of TO stars, so that no CO-depleted BSS is observed in our sample. 
The most intriguing result of this study concerns the BSS v $\sin(i)$ distribution which is shown in Fig. \ref{rotm4}:
most of the BSSs have low rotational velocities ($\lesssim$ 20 \kms) in agreement with the TO stars, but 8 out 20 BSSs have 
v $\sin(i) > $ 50 \kms. Only lower limits (marked with an arrow) have been computed for 6 of them, suggesting
v $\sin(i)$ higher than 70 \kms. This is the largest fraction (40\%) of fast rotating BSSs ever found in any GC.
\section{$\omega$ Centauri} 
\begin{figure}[]
\resizebox{\hsize}{!}{\includegraphics[clip=true]{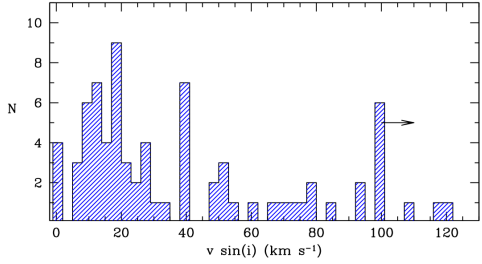}}
\caption{Rotational velocity distribution for BSSs in $\omega$ Centauri. Upper limits are marked with an arrow.
\footnotesize
}
\label{omegarot}
\end{figure}

$\omega$ Centauri is one of the most studied objects in the Milky Way since the 1960s.
All the evidence collected so far - kinematics, spatial distribution and chemical composition -
suggests that it is not a ``genuine'' GC but more likely the remnant of a dwarf galaxy that merged with the Milky Way in the past. 
The particular dynamical status of $\omega$ Centauri has been confirmed also by the radial distribution of its BSS population
that is not centrally peaked (as observed in all the dinamically evolved GCs) but completely flat, suggesting that
$\omega$ Centauri is not relaxed yet \citep{Ferraro06b, Ferraro12}. For this reason, 
the entire BSS population in $\omega$ Centauri has been suggested to be formed through MT in binary systems \citep{Ferraro06b}.\\
78 BSSs have been observed along the entire extension of the cluster. Their v $\sin(i)$ distribution is shown in Fig. 
\ref{omegarot}: most of the BSSs have values $<$ 20 \kms\ but a large 
fraction (corresponding to the 30\% of the sample) has v $\sin(i) > $ 50 \kms up to values larger than 100 \kms.
\section{M30}
\begin{figure}[]
\resizebox{\hsize}{!}{\includegraphics[clip=true]{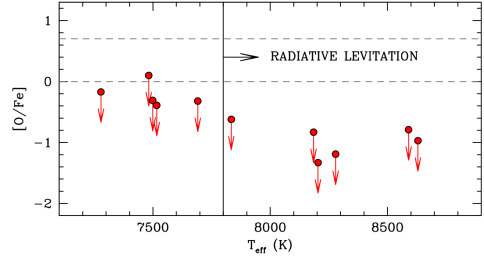}}
\caption{O abundances as a function of the effective temperature for the BSSs in M30 compared with RGBs values from \citet[][grey dashed lines]{Carretta09}.
The vertical line marks the limit for the occurrence of the radiative levitation.
\footnotesize
}
\label{m30ox}
\end{figure}

M30 is one of the 21 Galactic GCs that are likely to have experienced the core collapse 
\citep{Djorgovski86}. Two well distinct and almost 
parallel sequences of BSSs have been observed in the CMD of M30 by \citet{Ferraro09}. 
The suggested scenario is the following: 1-2 Gyr ago M30 underwent the core collapse 
that has boosted both the rate of direct stellar collisions, and the MT processes in 
binary systems. As a result, two different BSS sequences are now observed, the blue one formed
by COL-BSSs and the red one by MT-BSSs. We observed 12 BSSs, 4 in the blue sequence and 8 in the red one. 
The rotational velocity distribution shows that most of the BSSs rotate slowly with values ranging between 0 and 25 \kms. 
Only one fast rotating BSSs (having v $\sin (i) >$ 90 \kms) has been identified.
Moreover, there are hints that the BSSs in the blue sequence rotate faster than those in the red one.\\ 
Whereas the low metallicity of the cluster prevent us from obtaining meaningful C abundances, 
we derived upper limits for the O abundances (see Fig. \ref{m30ox}). Due to the radiative levitation effects (that alter the 
surface chemical abundances of BSSs hotter than $\sim$7800-8000 K, see \citealt{Lovisi12}), 
reliable upper limits for the O abundance have been obtained only for the 5 coldest BSSs (that are not affected 
by radiative levitation). The upper limits for 4 of them are incompatible with the O distribution of the giant stars 
in M30 \citep{Carretta09}, pointing out the occurence of an O-depletion among these BSSs. 
\section{Conclusions}
Chemical and kinematical properties of BSSs in three GCs have been studied. We found that BSSs in M4 do not show any evidence
of CO-depletion. This might be due to the fact that all the BSSs in this cluster have a collisional origin. Nevertheless, the size
of the sample is small and the lack of CO-depleted BSSs could be a statistical effect. Moreover, an intriguing scenario proposed by F06
suggests that the depletion is a transient phenomenon, since it could be erased by some mixing mechanism. We found evidence of O depletion in 4 out of the 
5 coldest BSSs of M30 that do not suffer from radiative levitation effects. In particular, we note that the O-depleted BSSs in M30 all 
belong to the red sequence that is suggested to be have a MT origin.\\
Concerning the rotational velocities, the distribution of the BSSs in M30 is very similar to that found in 47 Tuc by F06: almost all the BSSs rotate
slowly and only one BSS shows v $\sin(i) >$ 50 \kms. On the contrary, a large fraction of fast rotating BSSs (v $\sin(i) > $50 \kms) has been identified 
in M4 and $\omega$ Centauri, corresponding to the 40\% and 30\% of the entire sample respectively. These are the largest fractions of fast 
rotating BSSs ever found in any GC.

\begin{acknowledgements}
This research is part of the project COSMIC-LAB funded by the
European Research Council (under contract ERC-2010-AdG-267675).
\end{acknowledgements}

\bibliographystyle{aa}

\end{document}